\documentclass[fleqn,twoside,twocolumn,nofootinbib]{revtex4} 
\usepackage[sec]{ujp} 
\begin{document}
\title[THE STRUCTURE OF A CHAOS OF STRANGE ATTRACTORS WITHIN A MATHEMATICAL MODEL OF THE METABOLISM OF A CELL]
{THE STRUCTURE OF A CHAOS OF STRANGE ATTRACTORS WITHIN A MATHEMATICAL MODEL OF THE METABOLISM OF A CELL}%
\author{V.I.~Grytsay}
\affiliation{Bogolyubov Institute for Theoretical Physics,\\ Nat. Acad. of Sci. of Ukraine}
\address{14b, Metrolohichna Str., Kiev 03680, Ukraine}
\email{vgrytsay@bitp.kiev.ua}
\author{I.V.~Musatenko}
\affiliation{Kyiv National Taras Shevchenko University, Faculty of Cybernetics, Department of Computational Mathematics}%
\address{64, Vladimirskaya Str., 01033, Kyiv, Ukraine}%
\email{ivmusatenko@gmail.com}

\udk{}  \razd{\secx}

\setcounter{page}{1189}%
\maketitle

\begin{abstract}
This work continues the study of the earlier constructed mathematical model of the metabolic process running in a cell.

We will consider auto-oscillations arising on the level of enzyme-substrate interactions in the nutrient and respiratory chains, which leads to the self-organization in autocatalysis of the integral metabolic process in cells. The auto-oscillations organize themselves in the total metabolic process of cells at autocatalysis.

The behavior of the phase-parametric characteristic under the high
dissipation of a kinetic membrane potentialis analyzed. All possible
oscillatory modes of the system and the scenario of formation and
destruction of regular and strange attractors are studied. The
bifurcations of the transitions "order-chaos", "chaos-order",
 "chaos-chaos" and "order-order" are calculated. The total spectra of
Lyapunov indices and the divergences for all types of attractors on
a part of the phase-parametric characteristic under consideration
are determined. For various types of strange attractors, their
Lyapunov dimensions, KS-entropies, and "predictability horizons"
are calculated. Some conclusions about the structure of the chaos of
strange attractors and its influence on the stability of the
metabolic process in a cell are drawn.
\end{abstract}

\section{Introduction}

The study of the processes of self-organization in alive cells is one of the most important physical problems. Its solution will allow one to answer many questions about the physical nature of life. The most suitable object of studies is the metabolic processes, in which a complicated auto-oscillatory dynamics is manifested. Such modes were found in the processes of photosynthesis and glycolysis, the variation of the concentration of calcium in a cell, oscillations in heart muscle, etc. \cite{1}-\cite{4}.

A complicated auto-oscillatory dynamics is also revealed in widely applied bacteria Arthrobacter globiformis. These bacteria are used in the decomposition of practically all hydrocarbons of oil, purification of waste waters, production of tannic extracts, neutralization of the toxic action of herbicides on plants, biotechnologies, medicine, etc. In laboratory studies, the researchers have found the unpredictable behavior of these bacteria. The intensity of their growth can vary without apparent reasons. The multistability of stationary states and the auto-oscillatory modes of growth are revealed.

Earlier at G.K. Skryabin Institute of Biochemistry and Physiology of Microorganisms of the RAS, the studies aimed at the development of a biotechnological process of transformation of steroids by immobilized cells Arthrobacter globiformis in a bioreactor were carried out. By the results of those studies, V.P. Gachok and V.I. Grytsay constructed and studied a mathematical model of this.process. The stationary modes obtained within the model corresponded completely to experimental characteristics and were published in several works jointly with experimenters \cite{5}-\cite{8}. On the basis of the model, the appearance of various auto-oscillatory modes in granules with immobilized cells was theoretically established \cite{9}.

The first experimental confirmation of this phenomenon manifesting itself in the given population of cells was obtained later by researchers at the Institute of Microbiology of the RAS. The reason for its appearance was unknown to experimenters. For the quantitative interpretation of such unusual kinetics, they used a modified base synthetic chemostatic model (SCM) \cite{10}. This model is based on some hypotheses about the presence of some functions of a certain form, which describe the inhibition and the inactivation of metabolic processes in cells, as well as about the storage of glucose in the polymeric form. Under such conditions, an oscillatory process arises in the metabolism of a cell.

In 2002 at D.I. Mendeleev Russian Chemical-Technological University, a PhD dissertation was successively defended \cite{11}, in which the author constructed some model of auto-oscillations arising in a population of the given cells and cited the works by V.P. Gachok and V.I. Grytsay. The article describing the model was published somewhat later \cite{12}. In the article, it was stated that the mathematical model has a general character and can be applied to the description of oscillations of the biomass of various cells during their cultivation. The metabolic process running in a cell was described with some conditional intermediates. Oscillations in a population of cells appear due to the self-regulation of their number owing to feedbacks with delay under the loss of viability and the lysis of a part of cells due to the action of certain internal factors.

The basic distinction of the model developed by us from the above mentioned one consists in that we consider a specific real metabolic process of transformation of steroids, rather than a conditional one. We study the dynamics of variations in the concentrations of real metabolites in a cell under the breaking of the stationary modes, which corresponds to the experiment. The purpose of our studies is the consideration of intracellular oscillations arising on the levels of substrate-enzyme interactions and the respiratory chain. These are intracellular oscillations in the metabolic process. By this example, the laws of self-organization of these oscillations and the appearance of chaos in the metabolic process in a cell are investigated. Just the manifestation of such oscillations in the external medium is registered in experiments. In the earlier above-mentioned models, namely these extracellular oscillations were studied, but the internal dynamics of the cell itself was not considered.

By using the proposed model, we study a cell as an object of synergetics and as a nonlinear open self-organizing system. This approach allows us to consider the structural-functional connections inside a cell and to comprehend the physical laws of its vital activity.

In \cite{13}-\cite{21}, the model was used in the numerical calculations of spatio-temporal dissipative and chaotic structures formed with the participation of immobilized cells Arthrobacter globiformis in granules of a bioreactor. The study showed that the oscillations observed in the external solution of a bioreactor are created due to intracellular oscillations in the metabolism. The diffusion instability changes only the form of spatio-temporal structures. Therefore, there appears the necessity to study the dynamics of the metabolic process itself in a cell in more details. The numerical calculations within the model involved a variation of the dissipation of a kinetic membrane potential. This allowed us to determine the intervals on the phase-parametric characteristic, where the periodic, quasiperiodic, and chaotic attractors, whose dimension depends on the dissipation of a kinetic membrane potential, are present. A regularity in the sequence of the appearances of attractors on a toroidal surface was revealed. The sections, where the chaos arises by Feigenbaum's scenario, and an intermittence of the first kind were determined. The Poincar$\acute{e}$ sections and maps were studied, and the strange attractors, whose structure is formed with the help of folds and funnels, were found. In what follows, we will carry out the further study of the dynamics of a metabolic process within the given model, as well as the regularities of scenarios of formation of strange attractors under a high dissipation of the kinetic membrane potential.

\section{Mathematical model and methods of its study}

A mathematical model of the given metabolic process is constructed with regard for the general scheme of the metabolic process in cells Arthrobacter globiformis at a transformation of steroids \cite{22}-\cite{24}:
\begin{subequations}
\label{eq:system}
\begin{equation}
 \frac{dG}{dt}=\frac{G_{0}}{N_{3} + G +\gamma _{2} \psi}-l_{1}V(E_{1})V(G)-\alpha_{3}G,\label{subeq:1}
\end{equation}
\begin{equation}
 \frac{dP}{dt}=l_{1}V(E_{1})V(G)-l_{2}V(E_{2})V(N)V(P)-\alpha_{4}P,\label{subeq:2}
\end{equation}
\begin{eqnarray}
\frac{dB}{dt}=l_{2}V(E_{2})V(N)V(P)-k_{1}V(\psi)V(B)-\alpha_{5}B,\label{subeq:3}
\end{eqnarray}
\begin{eqnarray}
\frac{dE_{1}}{dt}=&&E_{1_{0}}\frac{G^{2}}{\beta_{1}+G^{2}} \left ( 1- \frac{P+mN}{N_{1}+P+mN}\right )- \nonumber\\&& -l_{1}V(E_{1})V(G)+l_{4}V(e_{1})V(Q)- \alpha_{1}E_{1},\label{subeq:4}
\end{eqnarray}
\begin{eqnarray}
\frac{de_{1}}{dt}=-l_{4}V(e_{1})V(Q)+l_{1}V(E_{1})V(G)-\alpha_{1}e_{1},\label{subeq:5}
\end{eqnarray}
\begin{eqnarray}
\frac{dQ}{dt}=&& 6lV \left ( 2-Q \right )V(O_{2})V^{(1)}(\psi)-\nonumber\\&& -l_{6}V(e_{1})V(Q)-l_{7}V(Q)V(N),\label{subeq:6}
\end{eqnarray}
\begin{eqnarray}
\frac{dO_{2}}{dt}=&& \frac{O_{2_{0}}}{N_{5}+O_{2}}-lV \left ( 2- Q \right ) \times \nonumber\\&& \times V(O_{2})V^{(1)}(\psi)-\alpha_{7}O_{2},\label{subeq:7}
\end{eqnarray}
\begin{eqnarray}
\frac{dE_{2}}{dt}=&&E_{2_{0}}\frac{P^{2}}{\beta_{2}+P^{2}}\frac{N}{\beta+N} \left (  1-\frac{B}{N_{2}+B} \right ) - \nonumber\\&&  -l_{10}V(E_{2})V(N)V(P)-\alpha_{2}E_{2},\label{subeq:8}
\end{eqnarray}
\begin{eqnarray}
\frac{dN}{dt}=&& -l_{2}V(E_{2})V(N)V(P)-l_{7}V(Q)V(N)+ \nonumber\\&& +k_{2}V(B)\frac{\psi}{K_{10}+\psi}+\frac{N_{0}}{N_{4}+N}-\alpha_{6}N,\label{subeq:9}
\end{eqnarray}
\begin{eqnarray}
\frac{d\psi}{dt}=l_{5}V(E_{1})V(G)+l_{8}V(N)V(Q)-\alpha \psi .\label{subeq:10}
\end{eqnarray}
\end{subequations}
where, $V(X)=X\diagup \left ( 1+X \right )$; $V^{(1)}(\psi)=1\diagup \left ( 1+\psi^{2} \right )$; $V(X)$ is a function describing the adsorption of the enzyme in the region of local coupling; and  $V^{(1)}(\psi)$ is a function characterizing the influence of the kinetic membrane potential on a respiratory chain.

The variables in the equations are dimensionless \cite{5, 7}.

We take the following values of parameters of the system:
$l=l_{1}=k_{1}=0.2$; $l_{2}=l_{10}=0.27$; $l_{5}=0.6$; $l_{4}=l_{6}=0.5$; $l_{7}=1.2$; $l_{8}=2.4$; $k_{2}=1.5$; $E_{1_{0}}=3$; $\beta_{1}=2$; $N_{1}=0.03$; $m=2.5$; $\alpha=0.033$; $a_{1}=0.007$; $\alpha_{1}=0.0068$; $E_{2_{0}}=1.2$; $\beta=0.01$; $\beta_{2}=1$; $N_{2}=0.03$; $\alpha_{2}=0.02$; $G_{0}=0.019$; $N_{3}=2$; $\gamma_{2}=0.2$; $\alpha_{5}=0.014$; $\alpha_{3}=\alpha_{4}=\alpha_{6}=\alpha_{7}=0.001$; $O_{2_{0}}=0.015$; $N_{5}=0.1$; $N_{0}=0.003$; $N_{4}=1$; $K_{10}=0.7$.

Equations~(\ref{subeq:1} - \ref{subeq:10}) describe variables of the concentrations: Eq.~(\ref{subeq:1}) -- hydrocortisone $(G)$; Eq.~(\ref{subeq:2}) -- prednisolone $(P)$; Eq.~(\ref{subeq:3}) -- $20\beta$-oxyderivative of prednisolone $(B)$; Eq.~(\ref{subeq:4}) -- oxidized form of 3-ketosteroid-- $\bigtriangleup '$ --dehydrogenase $(E_{1})$; Eq.~(\ref{subeq:5}) -- reduced form of 3-ketosteroid -- $\bigtriangleup '$-dehydrogenase $(e_{1})$; Eq.~(\ref{subeq:6}) -- oxidized form of the respiratory chain $(Q)$; Eq.~(\ref{subeq:7}) -- oxygen $(O_{2})$; Eq.~(\ref{subeq:8}) -- $20\beta$ -- oxysteroid-dehydrogenase $(E_{2})$; Eq.~(\ref{subeq:9}) -- $NAD \cdot H$  (reduced form of nicotinamide adenine dinucleotide) $(N)$.  Eq.~(\ref{subeq:10}) describes the variation of the kinetic membrane potential $(\psi)$.

The calculations are based on the Runge-Kutta-Merson method. The set accuracy is $10^{-8}$. Prior to the approach to an attractor by the system, the duration of the transient initial phase was taken to be $1.000.000$ in order to obtain the proper calculated values.

To construct the phase-parametric characteristic, we used the method of sections. In the phase space with a trajectory of the system, we drew the cutting plane $P=0.2$. Such a choice is supported by the symmetry of oscillations relative to this point in multiple modes.

The spectrum of Lyapunov indices was calculated, by using Benettin's algorithm with orthogonalization of the perturbation vectors within the Gram-Schmidt method \cite{23}.

Here, we consider the zero Lyapunov index to be a number, whose first significant number appears only in the fifth decimal place. By this, we identified the type of regular and strange attractors.

To classify the geometric structures of strange attractors, we calculated their fractal dimensions. Strange attractors are fractal sets and possess the Hausdorff-Besicovitch fractional dimension. But its direct calculation is a very difficult problem, which has no standard algorithm. Therefore, we calculated the Lyapunov dimension of attractors, as a quantitative measure of fractality, by the Kaplan-Yorke formula \cite{25, 26}
\begin{equation}
D_{\rm Fr}=m+\frac{ \sum ^{m}_{i=1} \lambda_{i}}{| \lambda _{m+1}|}
,\label{eq:2}
\end{equation}
where $m$ -- number of the first Lyapunov indices in the decreasing
order, whose sum $\sum^{m}_{i=1} \lambda _{i} \geq 0$; $m+l$ -
number of the first Lyapunov index, whose value $\lambda _{m+1}<0$.

In addition, we studied the variation of a distance between close
phase points of trajectories $d(t)=|x_{2}(t)-x_{1}(t)|$ during the
evolution of system (\ref{subeq:1}-\ref{subeq:10}). If the dynamics
of the system is chaotic, then $d(t)$ increases exponentially with
time: $d(t) \approx d(0) e^{kt}$. In this case, the mean rate of
divergence of trajectories is defined as $k=
\frac{ln[d(t)/d(0)]}{t}$. We consider also
\[h=
\lim_{d(0)\rightarrow 0 ~t\rightarrow \infty} \frac{\ln
[\frac{d(t)}{d(0)} ]}{t},\] which is called the Kolmogorov-Sinai
entropy or KS-entropy \cite{27, 28}. With the use of the KS-entropy,
we determined the conditions, under which the modes under study are
chaotic or regular attractors. In particular, if the dynamics of the
system is periodic or quasiperiodic, then the distance $d(t)$ does
not increase with time, and the KS-entropy is equal to zero $(h=0)$.
In the presence of a fixed point in the system, $d(t)\rightarrow 0$
and $h<0$. In the case of the chaotic dynamics of the system, the
KS-entropy is positive $(h>0)$.

Since the values of characteristic Lyapunov indices determine the rates of divergence of trajectories in the $m$-dimensional phase space of the system, we use the spectrum of Lyapunov indices for the calculation of the value of $h$. By the Pesin theorem \cite{29}, the KS-entropy corresponds to the sum of all positive Lyapunov characteristic indices:
\begin{equation}
h=\sum^{m}_{i=1}\lambda _{i}.\label{eq:3}
\end{equation}

The KS-entropy allows us to estimate the rate of loss of the information about the initial state of the system. The positiveness of the entropy is a criterion of the chaos. This gives the possibility to qualitatively evaluate properties of the local stability of attractors.

The value reciprocal to the KS-entropy,
\begin{equation}
t_{\min}=h^{-1},\label{eq:4}
\end{equation}
determines the time of mixing in the system and characterizes how
rapidly the initial conditions will be forgotten. At $t\ll
t_{\min}$, the behavior of the system can be predicted with
sufficient accuracy. At $t>t_{\min}$, only the probabilistic
description is possible. The chaotic mode is unforeseen due to the
loss of the memory of initial conditions. The quantity $t_{\min}$ is
called the Lyapunov index and characterizes the "predictability
horizon" of a strange attractor.

\section{Results of studies}

Earlier,  we studied the part of the phase-parametric characteristic
$\alpha \in (0.032,0.32554)$ and established that, at $\alpha =
0.032554$, regular attractor $10\cdot 2^{0}$ on a torus holds in the
system. We now continue the study of oscillatory modes of the system
under the variable dissipation of a kinetic membrane potential.
Below in (Table ~\ref{tab:tab1}), we present the calculated total
spectra of Lyapunov indices and the divergences for the majority of
modes under consideration.

\begin{figure*}
\includegraphics[width=15cm]{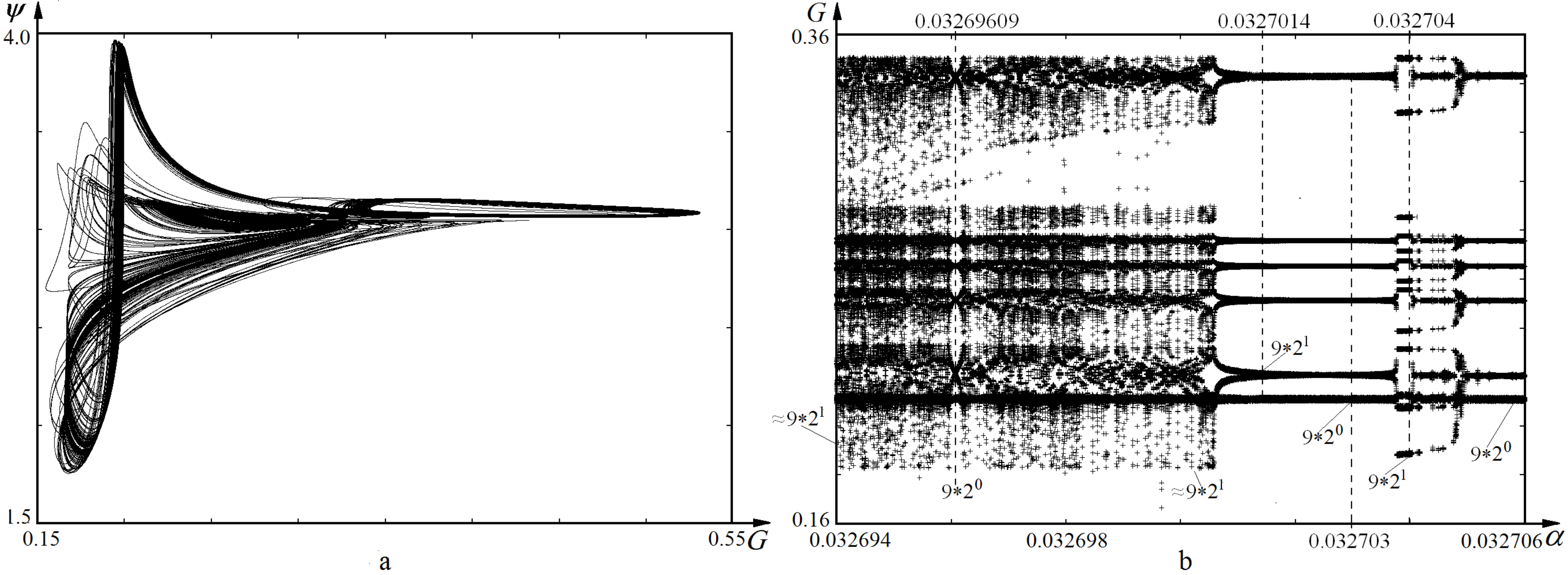}
\caption{\label{fig:pic1} a)  Projection of the phase portrait of
strange attractor $9\cdot2^{x}$ at $\alpha = 0.03269$; b)
phase-parametric characteristic of the system for $\alpha \in
(0.032694, 0.032706)$.}
\end{figure*}

Let  $\alpha$ increase. We observe that, at $\alpha = 0.0326735$,
strange attractor $10\cdot 2^{x}$ appears instantly. Hence, we
have the transition $10\cdot 2^{0} \rightarrow 10 \cdot 2^{x}$
of the "order-chaos" type. If the value of $\alpha$ grows, the
given strange attractor transits gradually to strange attractor
$9\cdot 2^{x} (\alpha = 0.03269)$ (Fig.~\ref{fig:pic1},a). In
this case, we observe the transition "chaos-chaos". Then strange
attractor $9\cdot 2^{x}$ shrinks to quasiperiodic cycle
$\approx 9\cdot 2^{1}$ on a torus $(\alpha=0.032694)$.

Let  us consider the part of the phase-parametric characteristic for
$\alpha \in (0.032694,0.032706)$ (Fig.~\ref{fig:pic1},b). By passing
from right to left, we see that regular attractor $9\cdot2^{0}$ on a
torus exists at $\alpha =0.032706$. As the dissipation of a kinetic
membrane potential in the interval $\alpha \in (0.032705,0.0327036)$
decreases, the oscillations on a toroidal surface are destroyed, and
the formation of the simple regular attractor $9\cdot 2^{1}$ with
doubled period occurs $(\alpha =0.032704)$. The further decrease in
$\alpha$ causes the renewal of cycle $9\cdot 2^{0}$ on a torus
$(\alpha = 0.032703)$. At $\alpha=0.0327014$, we observe the
formation of attractor $9\cdot 2^{1}$ and the second appearance of
the period doubling bifurcation on a torus. Then this cycle loses
the stability, and quasiperiodic cycle $\approx 9\cdot 2^{1}$ is
formed on a torodoidal surface $(\alpha =0.032697)$.

The  change of a section of the attractor of quasiperiodic cycle
$\approx9 \cdot 2^{1}$ in the interval $\alpha \in
(0.0327,0.032694)$ is shown in Fig. ~\ref{fig:pic1},b. At $\alpha =
0.03269609$, we see the sudden appearance of regular attractor
$9\cdot 2^{0}$ on a torus.

\begin{figure*}
\includegraphics[width=15cm]{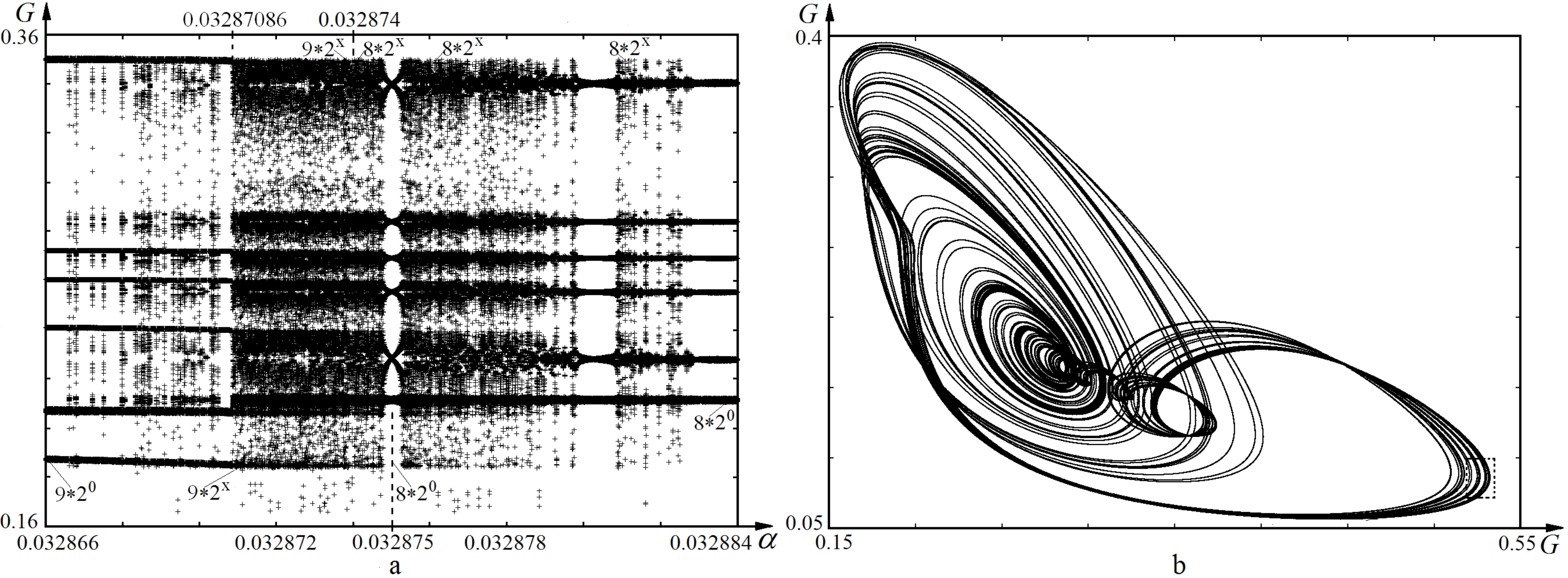}
\caption{\label{fig:pic2} a)  Phase-parametric characteristic of the
system for $\alpha \in (0.032866, 0.032884)$; b) projection of the
phase portrait of a strange attractor at the point of the mutual
transition "chaos-chaos": $9\cdot2^{x}\leftrightarrow
8\cdot2^{x}$ $(\alpha=0.032874)$.}
\end{figure*}

As $\alpha$  increases from $0.032706$ (Fig.~\ref{fig:pic1},b) to
$0.032866$ (Fig.~\ref{fig:pic2},a), the oscillations on a toroidal
surface of attractor $9\cdot 2^{0}$ cease, and the ordinary periodic
9-fold cycle is restored (Table ~\ref{tab:tab1}). But, the further
increase in $\alpha$ (Fig.~\ref{fig:pic2},a) leads to its aperiodic
fracture and to the formation of strange attractors $9\cdot
2^{x} (\alpha = 0.03287086)$. On the phase-parametric
characteristic (Fig.~\ref{fig:pic2},a), we observe the formation of
the zones of stability and instability of a regular attractor with
the appearance of the mutual transitions "order-chaos-order":
$9\cdot 2^{0} \leftrightarrow 9 \cdot 2^{x} \leftrightarrow 9
\cdot 2^{0}$. The further increase in $\alpha$ causes the
instability of strange attractors $9\cdot 2^{x}$ and their
self-organization in strange attractors $8 \cdot 2^{x}$.

In Fig.~\ref{fig:pic2},b, we show the projection of the phase
portrait of a strange attractor at the point of the mutual
transition "chaos-chaos": $9\cdot 2^{x} \leftrightarrow 8
\cdot 2^{x} (\alpha =0.032874)$. At $\alpha=0.032875$, there
occur the contraction of phase trajectories and the appearance of
regular attractor $8\cdot 2^{0}$ on a torus. As $\alpha$ increases,
the given regular attractor becomes unstable, and strange attractors
$8\cdot 2^{x}$ ($\alpha =0.0328765$, Table ~\ref{tab:tab1}) are
formed. The further growth of $\alpha$ causes their aperiodic
destruction and their alternation with regular attractors $8\cdot
2^{0}$ and $8\cdot 2^{1}$ on a torus. In the interval $\alpha \in
(0.03287980,0.0328808)$ (Fig.~\ref{fig:pic2},a), regular attractors
$8\cdot 2^{0}$ are conserved. Then, in the interval $\alpha \in
(0.0328809,0.0328828)$, strange attractors $8\cdot 2^{x}$
alternating with regular attractors of the 8-fold period are formed.
The transitions at the onset and the end of this interval, $8\cdot
2^{x} \leftrightarrow 8\cdot 2^{0} \leftrightarrow 8\cdot
2^{x}$ in both directions, are realized through the period
doubling bifurcation and the intermittence. Then, at $\alpha
=0.032884$, attractor $8\cdot 2^{0}$ arises again on a torus and
conserves its stability up to $\alpha =0.033117$. Then it contracts
in an ordinary periodic 8-fold cycle $(\alpha =0.0331)$
(Table~\ref{tab:tab1}).

\begin{figure*}
\includegraphics[width=15cm]{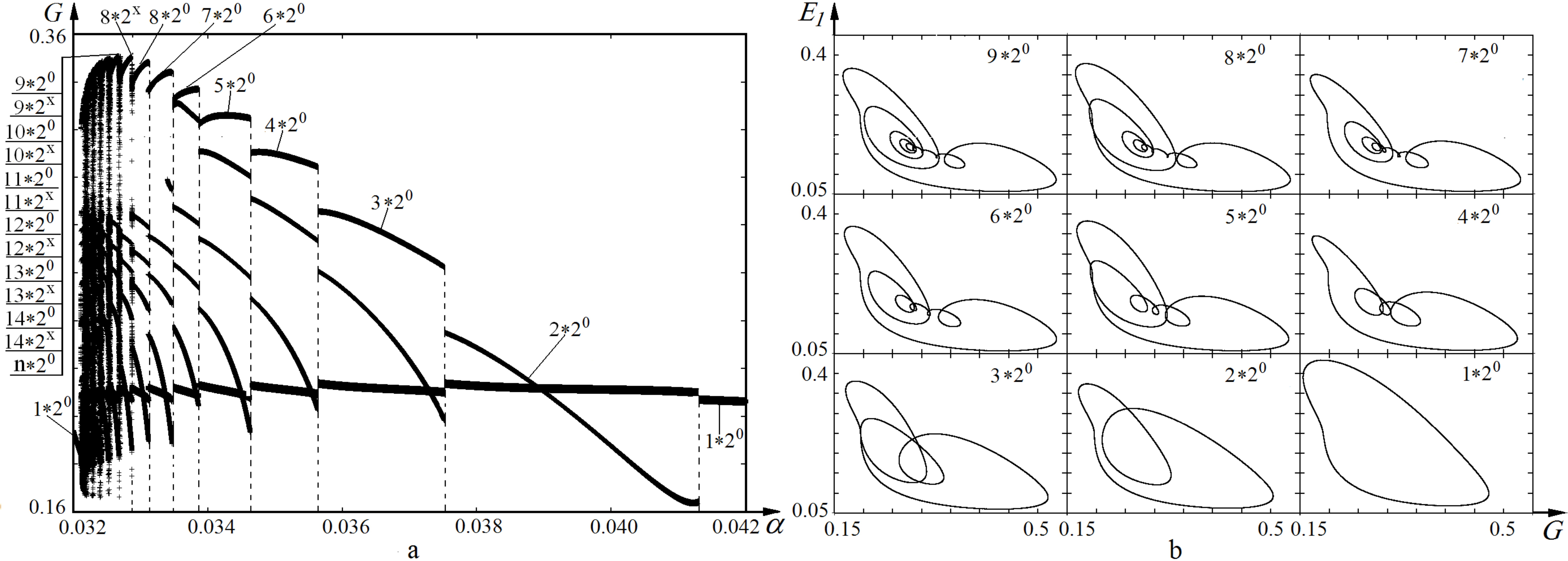}
\caption{\label{fig:pic3} a) Phase-parametric characteristic of the system for $\alpha \in (0.032, 0.042)$; b) projections of the phase portraits of regular attractors: $1\cdot2^{0}$ $(\alpha=0.042)$; $2\cdot2^{0}$ $(\alpha=0.039)$; $3\cdot 2^{0}$ $(\alpha=0.0375)$; $4\cdot 2^{0}$ $(\alpha=0.0348)$; $5\cdot 2^{0}$ $(\alpha=0.0346)$; $6\cdot 2^{0}$ $(\alpha=0.0338)$; $7\cdot 2^{0}$ $(\alpha=0.0332)$; $8\cdot 2^{0}$ $(\alpha=0.0331)$; $9\cdot 2^{0}$ $(\alpha=0.0328)$.}
\end{figure*}

The further variation of the multiplicity of an oscillatory process with increase in the dissipation of a kinetic membrane potential can be traced with the use of the whole phase-parametric characteristic (see Fig.~\ref{fig:pic3},a and Table~\ref{tab:tab1}).

Let  us consider Fig.~\ref{fig:pic3},a from right to left. At
$\alpha =0.042$, we observe a single periodic cycle $1\cdot 2^{0}$.
As $\alpha$ decreases, it is deformed, and its period increases due
to the bifurcation: $1\cdot 2^{0} \rightarrow 2\cdot 2^{0}$. The
further decrease in $\alpha$ causes a deformation of the new cycle
and the appearance of 3-fold regular attractor $3\cdot 2^{0}$ from
this cycle. Then the scenario is repeated. The appropriate cycle is
deformed, and its multiplicity increases by 1 at the points of
bifurcations. We see the successive increase in the multiplicity of
regular attractors. In Fig.~\ref{fig:pic3},b, we present the
projections for some examples of regular attractors formed in the
interval $\alpha \in(0.0328,0.042)$ with the following order of the
growing multiplicity of periods:
\begin{eqnarray*}
\begin{array}{l}
~~~~1\cdot2^{0}(\alpha=0.042)\rightarrow 2\cdot2^{0}(\alpha=0.039)\rightarrow \\
\rightarrow3\cdot2^{0}(\alpha=0.0375)\rightarrow 4\cdot2^{0} (\alpha=0.0348)\rightarrow \\
\rightarrow5\cdot2^{0} (\alpha =0.0346)\rightarrow6\cdot2^{0}(\alpha=0.0338)\rightarrow\\
\rightarrow7\cdot2^{0}(\alpha=0.0332)\rightarrow8\cdot2^{0}(\alpha=0.0331).
\end{array}
\end{eqnarray*}

As a characteristic specific feature of such transitions, we mention
the absence of strange attractors on the given part of the
phase-parametric characteristic. As $\alpha$ decreases, the
multiplicity of autooscillations in the metabolic process in a cell
varies discretely due to the the self-organization. We observe the
"order-order" transitions.

At the increase of $\alpha$ from $0.0331$ to $0.032884$ (Fig.~\ref{fig:pic2},a), regular attractor $8\cdot2^{0}$ holds.

In addition, Fig.~\ref{fig:pic3},a demonstrates a change of the dynamics of the metabolic process. In works \cite{22}-\cite{23}, this part of the phase-parametric characteristic was studied in detail, and the following scenario of variation of the multiplicity of attractors was found:
\begin{eqnarray*}
\begin{array}{l}
~~~~8\cdot2^{0}(\alpha=0.0331)\rightarrow 8\cdot2^{x}(\alpha=0.0328812)\rightarrow\\
\rightarrow9\cdot2^{x} (\alpha=0.03287086)\rightarrow 9\cdot2^{0} (\alpha=0.032866)\rightarrow \\
\rightarrow9\cdot2^{x}(\alpha =0.03269)\rightarrow10\cdot2^{x}(\alpha=0.0326735)\rightarrow\\
\rightarrow10\cdot2^{0}(\alpha=0.032554)\rightarrow10\cdot2^{x}(\alpha=0.03254)\rightarrow\\
\rightarrow11\cdot2^{x}(\alpha =0.032517)\rightarrow11\cdot2^{0}(\alpha=0.0325)\rightarrow\\
\rightarrow11\cdot2^{x}(\alpha=0.0324)\rightarrow12\cdot2^{x}(\alpha=0.032387)\rightarrow\\
\rightarrow12\cdot2^{0}(\alpha =0.032386)\rightarrow12\cdot2^{x}(\alpha=0.03229)\rightarrow\\
\rightarrow13\cdot2^{x}(\alpha=0.03227575)\rightarrow13\cdot2^{0}(\alpha=0.032275)\rightarrow\\
\rightarrow13\cdot2^{x}(\alpha =0.03222)\rightarrow14\cdot2^{x}(\alpha=0.03217)\rightarrow\\
\rightarrow\approx n\cdot2^{0}(\alpha=0.03215962)\rightarrow14\cdot2^{0}(\alpha=0.0321596)\rightarrow\\
\rightarrow8\cdot2^{x}(\alpha=0.03211295)\rightarrow8\cdot2^{0}(\alpha=0.0321107)\rightarrow\\
\rightarrow1\cdot2^{0}(\alpha=0.032).
\end{array}
\end{eqnarray*}

The transitions between multiple modes occur through the
intermediate formation of strange attractors. Here, we observe the
"order-chaos-chaos-order" transitions. At the limiting points of
the interval covered by Fig.~\ref{fig:pic3}, a, regular attractors
$1\cdot 2^{0}$ are established for the minimum and maximum
dissipations. Inside the interval, we see the appearance of regular
and strange attractors with various multiplicities as a result of
bifurcations and the self-organization.

\begin{figure*}
\includegraphics[width=15cm]{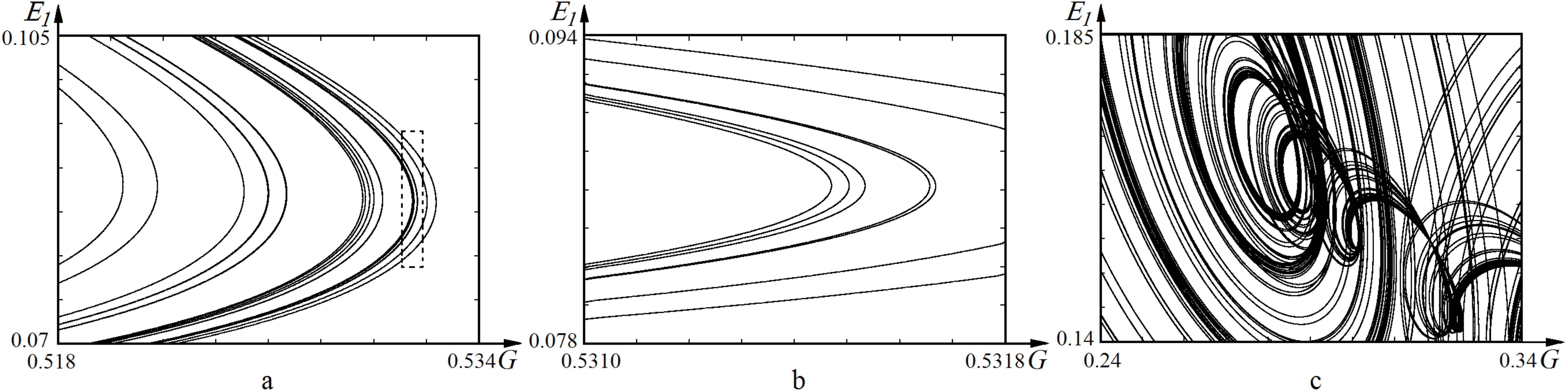}
\caption{\label{fig:pic4} Reconstruction of the fractal structure of strange attractor $(9\leftrightarrow 8)\cdot 2^{x}$ $(\alpha=0.032874)$; a) part separated by the dotted line in Fig.~\ref{fig:pic2},b; b - part separated by the dotted line in Fig.~\ref{fig:pic4},a; c) region of the formation of a deterministic chaos in the mixing funnel (see Fig.~\ref{fig:pic2},b) in the phase plane $(G,E_{1})$.}
\end{figure*}

We now consider the fractality of strange attractors. As an example, we take strange attractor $(9\leftrightarrow8)\cdot2^{x} (\alpha=0.032874)$ for the transition between 9- and 8-fold cycles. By separating a small rectangular region of the projection of the phase space in Fig.~\ref{fig:pic2},b, we map it onto Fig.~\ref{fig:pic4},a. We now separate a small rectangular region in Fig.~\ref{fig:pic4},a, which includes one of the phase curves and present it as Fig.~\ref{fig:pic4},b. As is seen, the character of the geometric structure of the given strange attractor is repeated on small and large scales of the projection of the phase portrait. Each arisen curve of the projection of a chaotic attractor is a source of formation of new curves. Moreover, the geometric regularity of a structure of trajectories in the phase space is repeated. In Fig.~\ref{fig:pic4},c, we show in detail that the geometric structure of the fractality is conserved also in the mixing funnel of the given strange attractor. This geometric structure reminds a two-scale Cantor parametric set.

Since the direct calculation of the fractal dimension of an
attractor is an extremely complicated problem, and no standard
algorithms are available for its solution, we restrict ourselves
only by the calculation of their Lyapunov dimension $D_{\rm Fr}$
Eq.(\ref{eq:2}).

We obtain that the Lyapunov dimension of all regular attractors of simple periodic cycles is equal to 1 and that of the regular attractors on a toroidal surface, which correspond to quasiperiodic cycles, is equal to 2.

We now calculate the fractal dimension for certain strange
attractors (see Table~\ref{tab:tab2}). In addition, we calculate
also the Kolmogorov--Sinai entropy $h$  Eq.(\ref{eq:3}) and the
Lyapunov index $t_{\min}$ Eq.(\ref{eq:4}).

\begin{figure*}
\includegraphics[width=15cm]{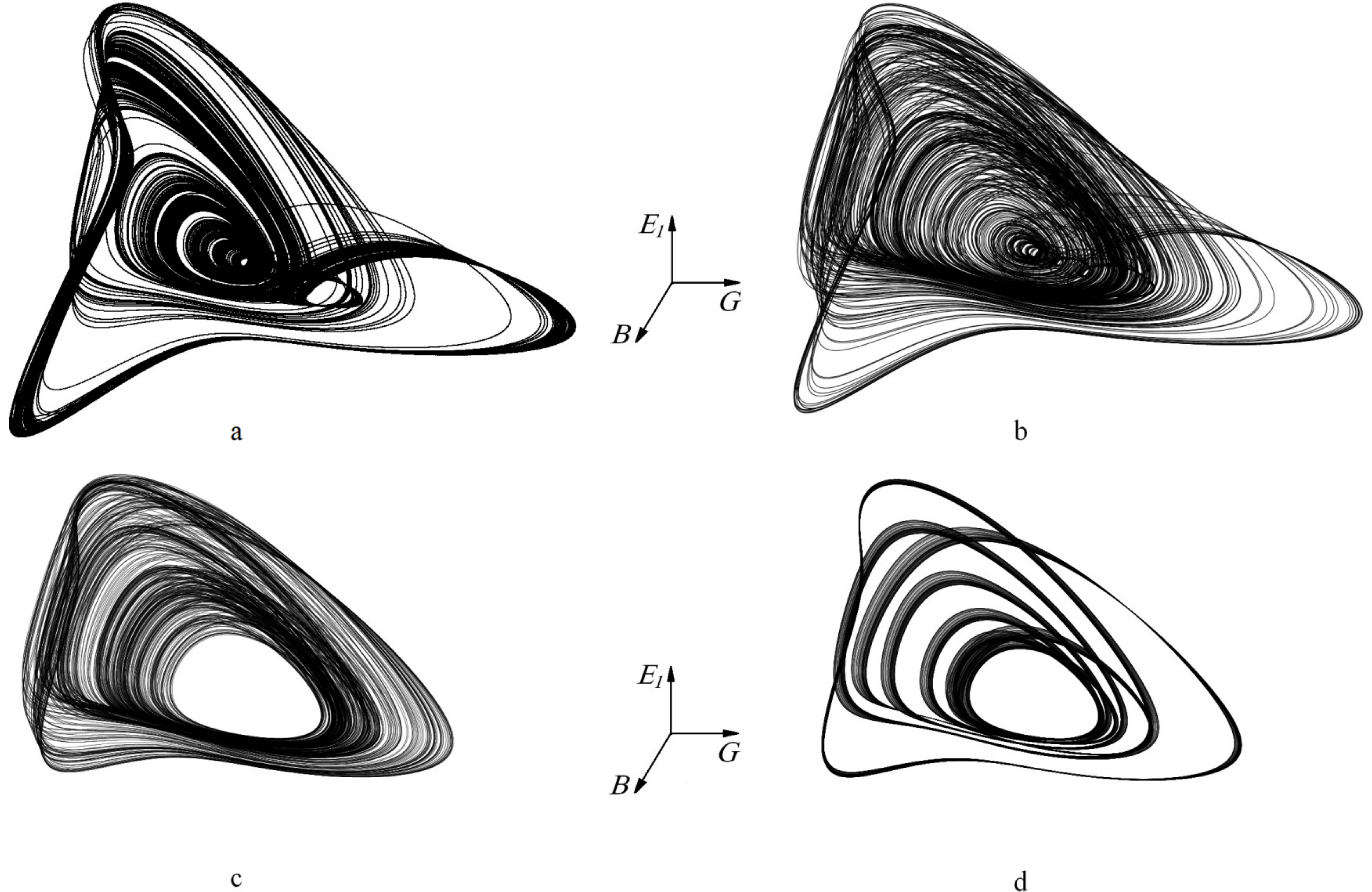}
\caption{\label{fig:pic5} Structure of the chaos of strange
attractors: a $8\cdot 2^{x}$ $(\alpha=0.0328715)$,
$t_{\min}=2597$, $h=0.000385$, $D_{\rm Fr}=2.073000$, $\Lambda =
-0.931419$; b $14\cdot2^{x}\leftrightarrow 7 \cdot 2^{x}$
$(\alpha=0.032168)$, $t_{\min}=1504$, $h=0.000665$, $D_{\rm
Fr}=2.216612$, $\Lambda = -0.901347$; d $8\cdot 2^{x}$
$(\alpha=0.0321195)$, $t_{\min}=3247$, $h=0.000308$, $D_{\rm
Fr}=2.150759$, $\Lambda = -0.898147$.}
\end{figure*}

Their phase portraits are presented in Fig.~\ref{fig:pic5}. At $\alpha=0.0328715$, strange attractor $8\cdot 2^{x}$ appears (Fig.~\ref{fig:pic5},a). Inside itself, it forms a funnel, where we observe the mixing of diverging trajectories of this unstable period. The deterministic chaos is forming. The further decrease of $\alpha$ is accompanied by a complication of the structure of the strange attractor. For example, at $\alpha=0.032168$ (Fig.~\ref{fig:pic5},b), a more complicated strange attractor arises as a result of the intermittence of two chaotic processes: $14\cdot 2^{x}$ and $7\cdot2^{x}$. Inside it, a funnel is located. The attractor is formed by two sources of chaotic autooscillatory processes: the transformation of steroids $G\rightarrow P\rightarrow B$ and a change of the activity of a respiratory chain due to variations of the kinetic membrane potential $\psi$. According to the definition given by Pomeau and Manneville, such a transition is called an intermittence of the first kind \cite{30}-\cite{33}.

At $\alpha=0.0321646$ (Fig.~\ref{fig:pic5},c), strange attractor $7\cdot2^{x}$ appears. Its specific feature is the absence of a funnel. Nevertheless, the attraction regions of an unstable 7-fold cycle are clearly distinguished on the phase portrait.

At $\alpha=0.03211295$ (Fig.~\ref{fig:pic5},d), we see strange attractor $8\cdot2^{x}$. It differs significantly from strange attractor $8\cdot2^{x}$ (Fig.~\ref{fig:pic5},a), has also no funnel, and the attraction regions of its trajectories are closed narrow strips with 8-fold period.

In this case, the KS-entropies of these modes decrease in the following sequence: $0.000665$ (Fig.~\ref{fig:pic5},b); $0.000437$ (Fig.~\ref{fig:pic5},c ), $0.000385$ (Fig.~\ref{fig:pic5},a), and $0.000308$ (Fig.~\ref{fig:pic5},d). The KS-entropy indicates the value of unpredictability for the motion of the phase trajectory of a strange attractor and characterizes the value of chaoticity of its deterministic chaos. The higher the KS-entropy, the greater the exponential divergence of phase trajectories along the perturbation vector corresponding to $\lambda_{1}$ and $\lambda_{2}$. For the rest vectors corresponding to the negative values of $\lambda_{3}-\lambda_{10}$ (Table~\ref{tab:tab2}), the phase trajectories exponentially contract to the own attractor. As the KS-entropy increases, the structure of a chaos is complicated (compare Fig.~\ref{fig:pic5},d,a,c,b). Among the modes under study, the mode shown in Fig.~\ref{fig:pic5},d is the most ordered.

In the same figures, we show the values of "predictability
horizon" and the fractal dimension of strange attractors. The mode
shown in Fig.~\ref{fig:pic5},d $(t_{\min}=3247)$ turns out to be the
most predictable as compared with those shown in
Fig.~\ref{fig:pic5},a $(t_{\min}=2597)$, Fig.~\ref{fig:pic5},c
$(t_{\min}=2288)$, and Fig.~\ref{fig:pic5},b $(t_{\min}=1504)$. In
these modes of deterministic chaos, the metabolic process is
predictable only in the determined time intervals $t_{\min}$.

As distinct from the KS-entropy, the Lyapunov dimension of the given
modes, which characterizes the fractality of these strange
attractors, increases by a somewhat different sequence:
Fig.~\ref{fig:pic5},a $(D_{\rm Fr}=2.073000)$, Fig.~\ref{fig:pic5},d
$(D_{\rm Fr}=2.150759)$, Fig.~\ref{fig:pic5},b $(D_{\rm
Fr}=2.216612)$, and Fig.~\ref{fig:pic5},c $(D_{\rm Fr}=2.323704)$.

This can be explained by that the Lyapunov dimension for these modes is determined not only by $\lambda_{1}$ and $\lambda_{2}$, but also by $|\lambda_{3}|$ Eq.(\ref{eq:2}), which characterizes the deformation of an element of the phase volume along the relevant perturbation vectors. The deformation increases with $\lambda_{1}$ and with decrease in $|\lambda_{3}|$.

In addition, the deformation of the total volume is also determined by the values of remaining negative Lyapunov indices $\lambda_{4}$ - $\lambda_{10}$ (Table~\ref{tab:tab2}). On the whole, the magnitude of such a change of the volume is determined by the divergence of a relevant mode (Table~\ref{tab:tab2}), namely: $\Lambda=-0.898147$ (Fig.~\ref{fig:pic5},d), $\Lambda=-0.901347$ (Fig.~\ref{fig:pic5},b), $\Lambda=-0.903255$ (Fig.~\ref{fig:pic5},c) and $\Lambda=-0.931419$ (Fig.~\ref{fig:pic5},a).

The mode in Fig.~\ref{fig:pic5},d is the most functionally stable
for a cell. The metabolic process in a cell is self-organized, so
that it possesses the least dissipation and the highest
"predictability horizon". At the given coefficient of dissipation
of the kinetic membrane potential, $\alpha=0.03211295$, the
insignificant variations of parameters of the system do not cause a
change of the dynamics of the metabolic process.

Analyzing the presented typical strange attractors, it is possible to indicate a certain regularity of the hierarchy of the chaotic behavior in the metabolism of a cell. The structures different in their characteristics replace one another. As a result, the cell adapts to varying conditions of the environment and conserves its functionality.

\begin{table*}
\caption{\label{tab:tab1}Total spectra of Lyapunov indices for attractors of the system under study (We do not show $\lambda_{4}$-$\lambda_{9}$, because they are not important for our conclusions).}\vskip3mm
\begin{ruledtabular}
\begin{tabular}{llccccccccccc}
$\alpha$ &~~~Attractor  & $\lambda_{1}$ & $\lambda_{2}$ & $\lambda_{3}$ & ... & $\lambda_{10}$ & $\Lambda$\\
\hline
.032554 & $~~~10\cdot 2^{0}(t)$ & ~.000024 & -.000001 & -.004607 & ... & -.509882 & -.918426  \\
0.0326735 & $~~~10\cdot 2^{x}$ & ~.000444 & -.000027 & -.005019 & ... & -.514817 & -.922863 \\
0.03269 & $~~~9\cdot 2^{x}$ & ~.000396 & ~.000035 & -.004854 & ... & -.516039 & -.924688 \\
0.032694 & $\approx 9\cdot 2^{1} (t)$ & ~.000054 & -.000029 & -.004812 & ... & -.515355 & -.925486\\
0.03269609 & $~~~9\cdot 2^{0}(t) $ & ~.000056 & ~.000018 & -.004805 & ... & -.514496 & -.924643\\
0.032697 & $\approx 9\cdot 2^{1}(t)$ & ~.000070 & ~.000004 & -.004804 & ... & -.514505 & -.924654\\
0.0327014 & $~~~9\cdot2^{1}(t)$ & ~.000031 & ~.000002 & -.004801 & ... & -.514363 & -.924584\\
0.032703 & $~~~9\cdot2^{0}(t)$ & ~.000023 & ~.000001 & -.004797 & ... & -.514369 & -.924627\\
0.032704 & $~~~9\cdot2^{1}$ & ~.000052 & -.000425 & -.004113 & ... & -.516335 & -.925214\\
0.032706 & $~~~9\cdot2^{0}(t)$ & ~.000022 & -.000019 & -.004790 & ... & -.514520 & -.924884\\
0.032866 & $~~~9\cdot2^{0}$ & ~.000049 & -.000851 & -.004304 & ... & -.516447 & -.927297\\
0.03287086 & $~~~9\cdot2^{x}$ & ~.000154 & -.000077 & -.005312 & ... & -.518125 & -.928356\\
0.032874 & $9\cdot2^{x}\leftrightarrow 8 \cdot2^{x}$ & ~.000111 & ~.000008 & -.005054 & ... & -.521442 & -.933349\\
0.032875 & $~~~8\cdot2^{0}(t)$ & ~.000044 & ~.000031 & -.005042 & ... & -.521542 & -.933617\\
0.0328765 & $~~~8\cdot2^{x}$ & ~.000364 & ~.000010 & -.005139 & ... & -.522725 & -.933812\\
0.032877 & $~~~8\cdot2^{1}(t)$ & ~.000042 & -.000034 & -.005031 & ... & -.521540 & -.933605\\
0.032884 & $~~~8\cdot2^{0}(t)$ & ~.000020 & ~.000013 & -.005039 & ... & -.521533 & -.933816\\
0.0331 & $~~~8\cdot2^{0}$ & ~.000041 & -.001030 & -.004499 & ... & -.522401 & -.936430\\
0.0332 & $~~~7\cdot2^{0}$ & ~.000013 & -.000362 & -.005280 & ... & -.530190 & -.946471\\
0.0338 & $~~~6\cdot2^{0}$ & ~.000031 & -.001004 & -.005479 & ... & -.539588 & -.961956\\
0.0346 & $~~~5\cdot2^{0}$ & ~.000044 & -.001560 & -.005593 & ... & -.552388 & -.980067\\
0.0348 & $~~~4\cdot2^{0}$ & -.000016 & -.001151 & -.006220 & ... & -.576517 & -1.008694\\
0.0375 & $~~~3\cdot2^{0}$ & ~.000024 & -.002174 & -.006569 & ... & -.595931 & -1.048803\\
0.039 & $~~~2\cdot2^{0}$ & ~.000004 & ~.002403 & -.006890 & ... & -.647123 & -1.112671\\
0.042 & $~~~1\cdot2^{0}$ & ~.000025 & -.001510 & -.007163 & ... & -.712164 & -1.194750\\
\end{tabular}
\end{ruledtabular}
\end{table*}

\begin{table*}
\caption{\label{tab:tab2}Total spectra of Lyapunov indices for strange attractors in Fig.~\ref{fig:pic5} (We do not show $\lambda_{4}$-$\lambda_{9}$, because they are not important for our conclusions).}\vskip3mm
\begin{ruledtabular}
\begin{tabular}{llccccccccccc}
$\alpha$ & ~~~Attractor  & $\lambda_{1}$ & $\lambda_{2}$ & $\lambda_{3}$ & ... & $\lambda_{10}$ & $\Lambda$\\
\hline
.0328715 & $~~~~~8\cdot 2^{x}$ & .000377 & .000008 & -.005274 & ... & -.520953 & -.931419\\
.032168 & $14\cdot 2^{x}\leftrightarrow 7\cdot 2^{x}$ & .000633 & .000032 & -.003070 & ... & -.504745 & -.901347\\
.0321646 & $~~~~~7\cdot 2^{x}$ & .000424 & .000013 & -.001350 & ... & -.515424 & -.903255\\
.03211295 & $~~~~~8\cdot 2^{x}$ & .000264 & .000044 & -.002043 & ... & -.507700 & -.898147\\
\end{tabular}
\end{ruledtabular}
\end{table*}

\section{Conclusions}

With the help of the mathematical model of a cell, we have performed
the study of the dynamics of the metabolic process in the mode of
oscillations under the enhanced dissipation of a kinetic membrane
potential. The scenarios of formation and destruction of regular and
strange attractors with various periods and types are determined.
The boundaries of the phase-parametric characteristics of regions,
where the bifurcations and the transitions "chaos-order",
 "order-chaos", "chaos-chaos", and "order-order" arise, are given.
The total spectra of Lyapunov indices and the divergences are
calculated. For some typical types of strange attractors, we have
determined Lyapunov dimensions of their fractality, KS-entropies,
and "predictability horizons". The structure of the chaos of
attractors, the hierarchy of their kinds, and the influence of the
chaos on the stability of the metabolic process, and the adaptation
and the functioning of a cell are studied.

One of the purposes  of the present work is to demonstrate a
possibility to apply the mathematical apparatus of nonlinear
dynamics to the study of the dynamics of metabolic processes within
a specific model. This allows us to consider the
structural-functional connections in a cell and the laws of its
self-organization. These systems are the excellent field for
applying the methods of nonlinear dynamics to the analysis of the
multidimensional systems of nonlinear differential equations.

The work is supported by the project N 0112U000056 of the National Academy of Scienses of Ukraine.

\vskip-5mm
\rezume{%
СТРУКТУРА ХАОСУ ДИВНИХ АТРАКТОРІВ\\ МАТЕМАТИЧНОЇ МОДЕЛІ МЕТАБОЛІЗМУ
КЛІТИНИ}{В.Й. Грицай, І.В. Мусатенко} {Дана робота є продовженням
досліджень побудованої раніше математичної моделі метаболічного
процесу клітини. Досліджуються автоколивання, що виникають на рівні
фермент-субстратных взаємодій і дихального ланцюга. При автокаталізі
відбувається їх самоорганізація в єдиному метаболічному процесі
клітини. Продовжуються дослідження фазопараметричної характеристики
при підвищенній дисипації кінетичного мембранного потенціалу.
Досліджено всі можливі коливальні режими системи. Вивчено сценарій
формування і руйнації регулярних та дивних атракторів. Знайдено
біфуркації переходів "порядок-хаос", "хаос-порядок",
"хаос-хаос" и "порядок-порядок". Знайдено повні спектри
показників Ляпунова і дивергенції для всіх видів атракторів на
розглядаємій ділянці фазопараметричної характеристики. Для різних
типів дивних атракторів розраховано їх ляпуновскі розмірності,
КС-ентропії та "горизонти передбачуваності". Зроблені висновки про
структуру хаосу дивних атракторів і його вплив на стійкість
метаболічного процесу клітини. }

\vskip-5mm
\rezume{%
СТРУКТУРА ХАОСА СТРАННЫХ АТТРАКТОРОВ МАТЕМАТИЧЕСКОЙ МОДЕЛИ МЕТАБОЛИЗМА КЛЕТКИ}
{В.И. Грицай, И.В. Мусатенко} {Данная работа является продолжением исследований построенной ранее математической модели метаболического процесса клетки. Исследуются автоколебания, возникающие на уровне фермент-субстратных взаемодействий и дыхательной цепи. При автокатализе происходит их самоорганизация в едином метаболическом процессе клетки. Продолжаются исследование фазопараметрической характеристики при повышенной диссипации кинетического мембранного потенциала. Исследованы все возможные колебательные режимы системы. Изучен сценарий формирования и разрушения регулярных и странных аттракторов. Найдены бифуркации переходов "порядок-хаос", "хаос-порядок", "хаос-хаос" и "порядок-порядок". Найдены полные спектры показателей Ляпунова и дивергенции для всех видов аттракторов на рассматриваемом участке фазопараметрической характеристики. Для различных типов странных аттракторов рассчитано их ляпуновские размерности, КС-энтропии и "горизонты предсказуемости". Сделаны выводы о структуре хаоса странных аттракторов и его влияние на устойчивость метаболического процесса клетки.}

\end{document}